\documentclass[slac_one]{revtex4}
\usepackage{graphicx}
\usepackage{fancyhdr}
\newcommand{\eh}[1]{\,\mathrm{#1}}
\newcommand{\dg}{^{\circ}}
\newcommand{\mr}[1]{\mathrm{#1}}
\newcommand{\prc}{\eh{\%}}
\newcommand{\lgt}{\log_{10}}
\newcommand{\astroph}{Astropart.~Phys.}

\begin{document}

\title{Cosmic Ray Results from the IceTop Air Shower Array} 

\author{Hermann Kolanoski (for the IceCube Collaboration$^*$)}
\affiliation{Institut f\"ur Physik, Humboldt-Universit\"at zu Berlin, D-12489 Berlin, Germany \\ 
$^*$http://icecube.wisc.edu}

\begin{abstract}
We report on the first results obtained with the IceTop air shower array on the cosmic ray energy spectrum and mass composition in the range of 1 PeV to 80 PeV. IceTop is the surface detector of the IceCube neutrino telescope currently under construction at the South Pole. 
A high sensitivity to the primary mass composition was observed by reconstructing showers at different zenith angles. 
Assuming only protons or iron nuclei as primary particles yields significantly different energy spectra for different zenith angle ranges, while only models with mixed composition, like the poly-gonato model, lead to the expected isotropic flux.
The prospects of composition measurements with different, alternative methods using the full IceCube detector are also discussed.
\end{abstract}

\maketitle

\thispagestyle{fancy}

\section{INTRODUCTION\label{sec_intro}} 
IceTop, the surface component of the IceCube Neutrino Telescope at the South Pole, is an air-shower array which will finally consist of 160 ice tanks for the detection of Cherenkov radiation at 80 stations spread over an area of 1\,km$^2$ \cite{icetop_icrc03} (Fig.\,\ref{fig_icecube}).  The detector is designed to measure air showers initiated by primary cosmic rays in  the energy range from about 300\,TeV up to about 1\,EeV, covering the `knee' region, where the observed cosmic ray energy spectrum changes spectral index, and possibly also including a transition from galactic to extra-galactic cosmic rays. In most models, the observed steepening of the spectrum is  expected to be accompanied by a change in the chemical composition of the primaries. Though there are experimental indications for such a change in composition, the details of the features of the energy spectrum in that region are not well known. The investigation of the energy spectrum and the composition in this energy range is therefore the prime goal of cosmic ray physics with IceCube. 

IceTop, together with the in-ice detector, which is located at a depth between 1450 and 2450 m in the ice, has quite unique properties for the determination of the cosmic ray composition. Most important for these studies is the possibility to measure the electromagnetic component of the air shower with the surface detector in coincidence  with high-energy muons in the deep ice. The ratio of the number of high-energy muons to the electromagnetic shower energy is the most sensitive parameter for the composition analysis.     

The first analysis of high-energy air showers with IceTop,
which we are presenting here, uses IceTop data alone. These data were taken within a single month with the 2007 detector configuration consisting of 26 stations. Due to the high altitude at the South Pole, with an overburden of only 700\,g/cm$^2$, the showers near the vertical are detected close to the shower maximum, yielding a strong zenith angle dependence due to varying slant depths. Scanning in this way the shower shape leads to a high sensitivity to composition.

\section{EXPERIMENT AND DETECTOR}
\begin{figure}
\begin{minipage}[b]{0.47\linewidth} 
\centering
\includegraphics[width=0.55\textwidth]{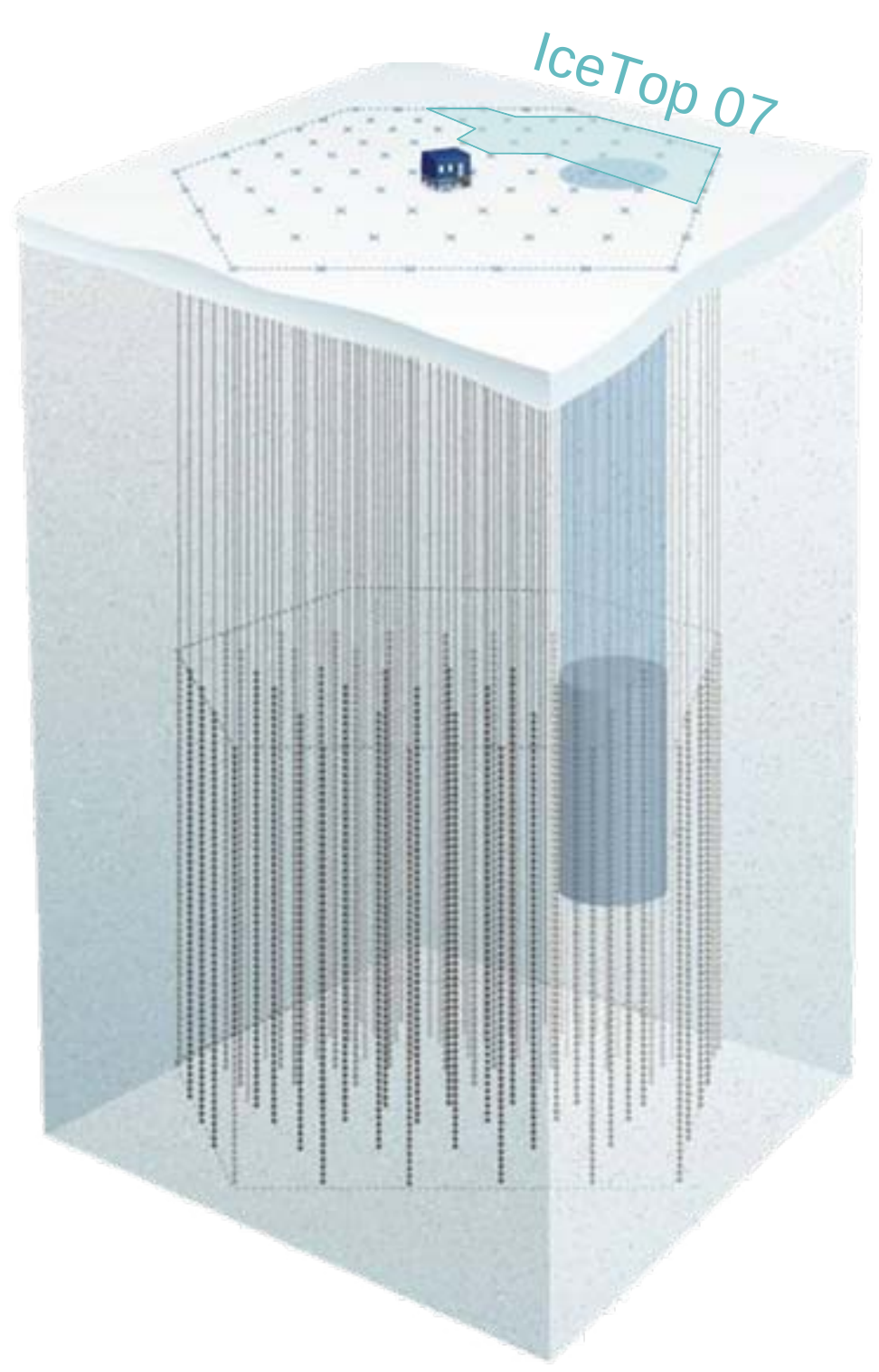}
\vspace{-5mm}
\caption{The IceCube Telescope with the 80 strings in the deep ice and 80 stations of the surface detector IceTop. Indicated are the IceTop 26 stations of the configuration in 2007 when the data described in this paper were taken.\label{fig_icecube}}
\end{minipage}
\hspace{0.5cm} 
\begin{minipage}[b]{0.47\linewidth}
\centering
\includegraphics[width=1.0\textwidth]{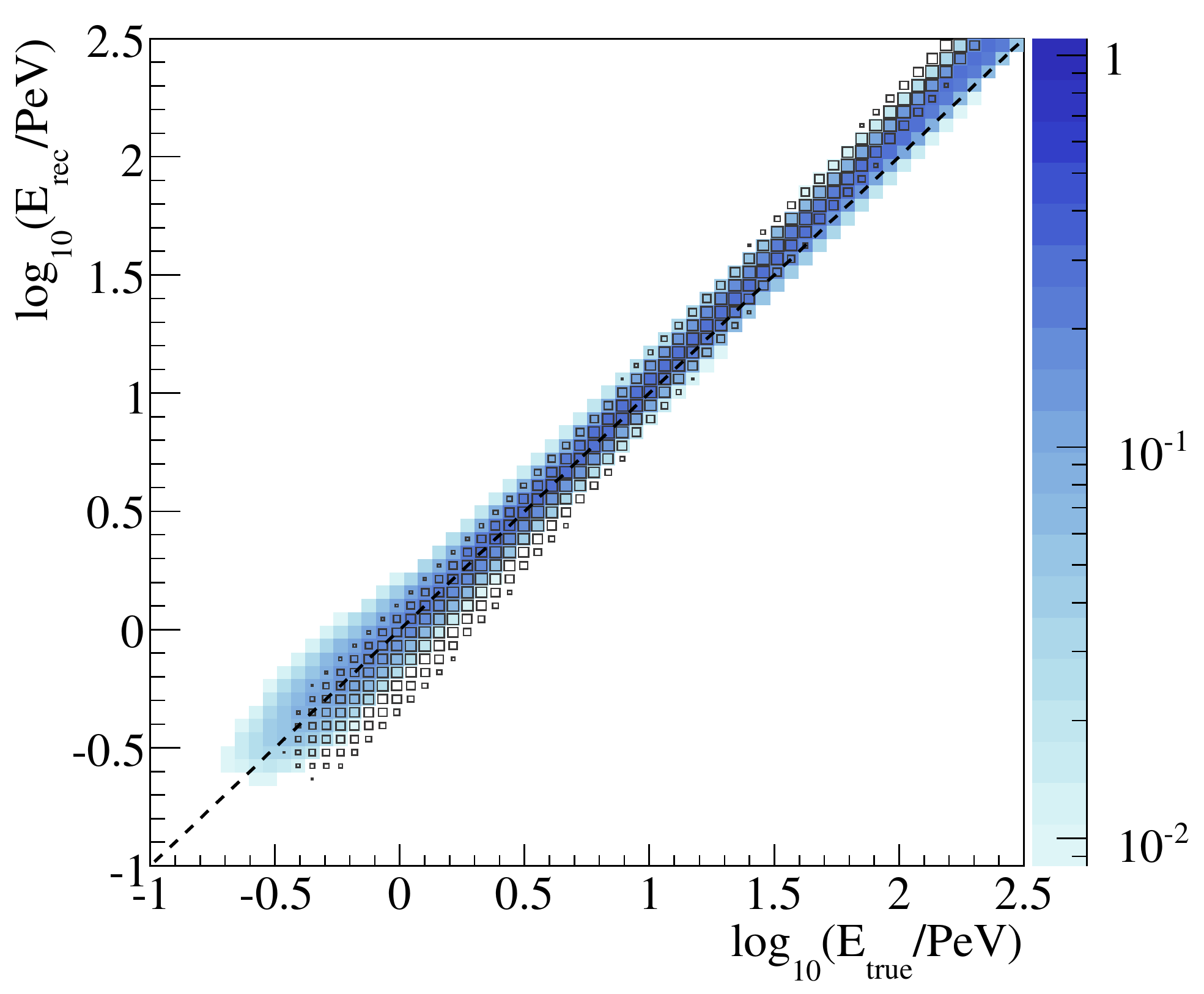}
\caption{Graphical display of the proton (blue/light) and iron (black) response matrices for air
showers below a zenith angle of $30\dg$.}
\label{response_matrix}
\end{minipage}
\end{figure}

The final IceTop detector will consist of 80 stations, each one located close to the top of an IceCube string,  distributed on a grid with mean distances of 125\,m (Fig.\,\ref{fig_icecube}). Each station comprises two cylindrical tanks with a diameter of $1.86\eh{m}$ filled with ice to a depth of $90\eh{cm}$. Each tank is equipped with two `Digital Optical Modules' (DOMs) containing a $10''$ photo multiplier tube to record the Cherenkov light of charged particles that penetrate the tank. In addition, a DOM houses complex electronic circuitry supplying signal digitisation, readout, triggering, calibration, data transfer and various control functions. 
The most important feature of the DOM electronics is the recording of the analog waveforms in $3.3\eh{ns}$ wide bins for a duration of 
$422\eh{ns}$.
The two DOMs in each tank are operated at different gains to enhance the dynamic range. 
The measured charges are expressed in units of `vertical equivalent muons'
(VEM) determined by calibrating each DOM with muons.

To initiate the readout of DOMs, a `local coincidence' of the two high gain DOMs of a station is required (threshold about 1/6 of a VEM). 
The data is written, and thus available for analysis, if the readouts of six or more DOMs are launched by a `local coincidence'. With 26 stations in 2007, the trigger rate was about $14\eh{Hz}$.

\section{AIR SHOWER RECONSTRUCTION}

For each triggered tank in an event, the time and charge of the signal are evaluated for further processing. (The full waveform is available but has not been exploited for this analysis). 
Likelihood maximisation methods are used to reconstruct the location,
direction and size of the recorded showers.  In general, the signal times
contain the direction information, and the charge distribution is connected to the shower size and core location. This analysis required five or more triggered stations leading to a reconstruction threshold of about $500\eh{TeV}$. A constant efficiency was reached at about $1\eh{PeV}$, depending on shower inclination.

Starting with some first guess values for the fit parameters location, direction and size of the shower, the parameters were determined either by a combined fit or iteratively by two separate fits, one for the direction and the other for the core location and shower size.  The shower direction was determined by fitting the sum of a paraboloid and a Gaussian (centered around the core position; $\sigma \approx120\eh{m}$) to the lateral distribution of signal times, yielding a resolution of $1.5\dg$, almost independent of energy and zenith angle.
The core position was determined by a  fit to the lateral signal distribution using the function 
\begin{equation}\label{eq:dlp}
          S(r) =
S_{\mathrm{ref}}\left(\frac{r}{R_{\mathrm{ref}}}\right)^{-\beta_{\mathrm{ref}} - \kappa\,
                   \log_{10}\left(\frac{r}{R_{\mathrm{ref}}}\right)}.
\end{equation}
where $r$ is the perpendicular distance to the shower axis (depending on the shower core position and direction), $S_{\mathrm{ref}}$ is the fitted signal at a distance
$R_{\mathrm{ref}}$, $\beta_{\mathrm{ref}}$ a slope parameter related to the shower age, and $\kappa$ a (lateral) curvature parameter. In the fit, $R_{\mathrm{ref}}=125\eh{m}$, the grid spacing, was used. The parameter $\kappa=0.303$ was found constant in simulations and remained fixed in
the fit.

The assignment of an energy to an air shower depends on the species of the primary nucleus. In first approximation only protons as primaries were assumed, and an energy estimator was obtained as a function of the fitted signal $S_{\mathrm{ref}}$ at the reference radius $R_{\mathrm{ref}}$ and  of the zenith angle $\theta$:
	$E_{\mathrm{rec}}=E(S_{\mathrm{ref}}, \theta)$.
The energy resolution improves with energy and approaches $0.05$ in $\lgt E$, or $12\prc$ in $E$, at $\sim 3\eh{PeV}$ for zenith angles below $30\dg$. 

To derive the true energy spectrum,  a response matrix $(M_{ij})$ was defined which relates the bin contents of the first guess energy distribution to the bin contents of the true energy spectrum: $N_i(E_{\mathrm{rec}}) = \sum_j M_{ij}\, N_j(E_{\mathrm{true}})$.  
Response matrices have to be determined for different primary mass compositions and zenith angles by detailed simulations of  the detector efficiency and energy smearing. The response matrices obtained for protons and iron nuclei are shown in Fig\,\ref{response_matrix}.
The faster development of showers from heavy primaries leads to a tilt of the iron band against the proton band (with the detailed simulation the proton band as well is not perfectly diagonal although the first guess assumes protons).
Since IceTop is close to the shower maximum, the center of rotation (where the two bands cross) lies within the observed energy range. This means that at low energies, showers from heavy primaries look less energetic
than proton showers, whereas at high energies they appear more energetic. It should be noted that the point of rotation depends on many factors, such as the chosen reference radius for the energy extraction and inclination.

\begin{figure}
\centerline{
{\includegraphics[width=0.42\columnwidth]{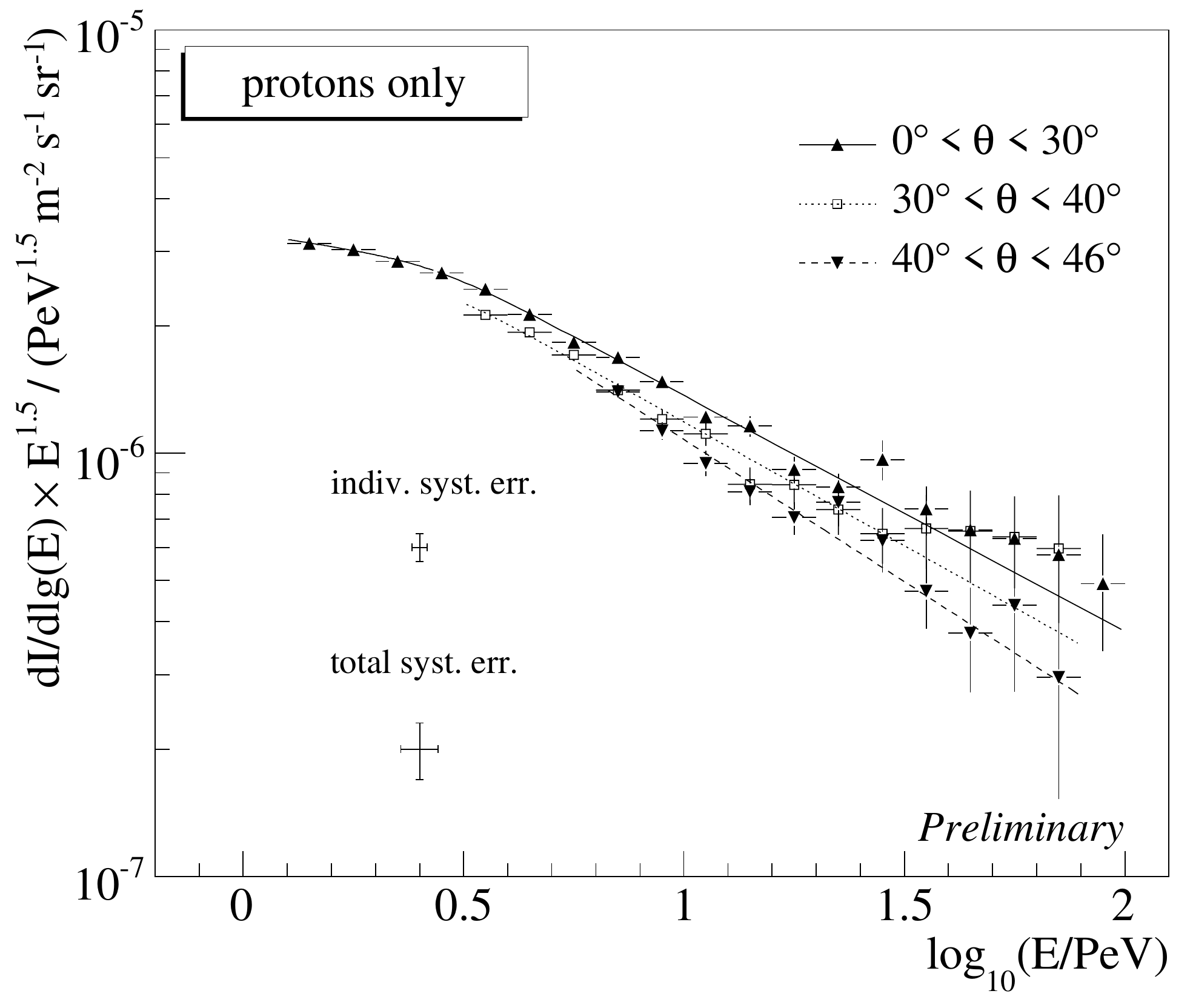}
\label{proton}}
\hfil
{\includegraphics[width=0.42\columnwidth]{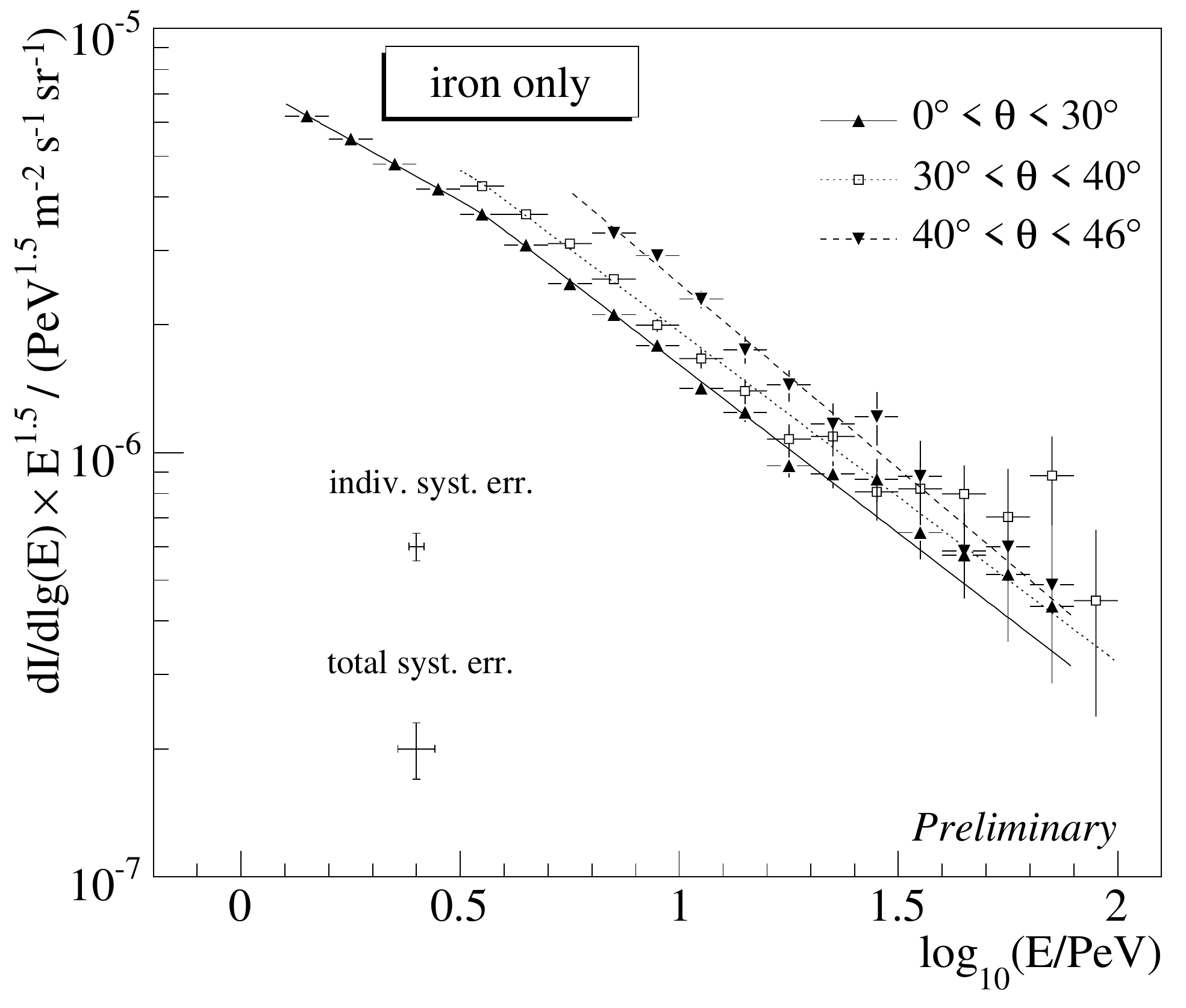}
\label{iron}}}
\vspace{-2mm}
\caption{Preliminary, unfolded energy spectra for three zenith bands, assuming pure proton (left) and pure iron compositions (right). 
The error bars of the single points represent statistical errors. The total systematic error, and the error intrinsic to the inclination bins, are displayed on the lower left. \label{unfolding_pFe}}
\centerline{
{\includegraphics[width=0.42\columnwidth]{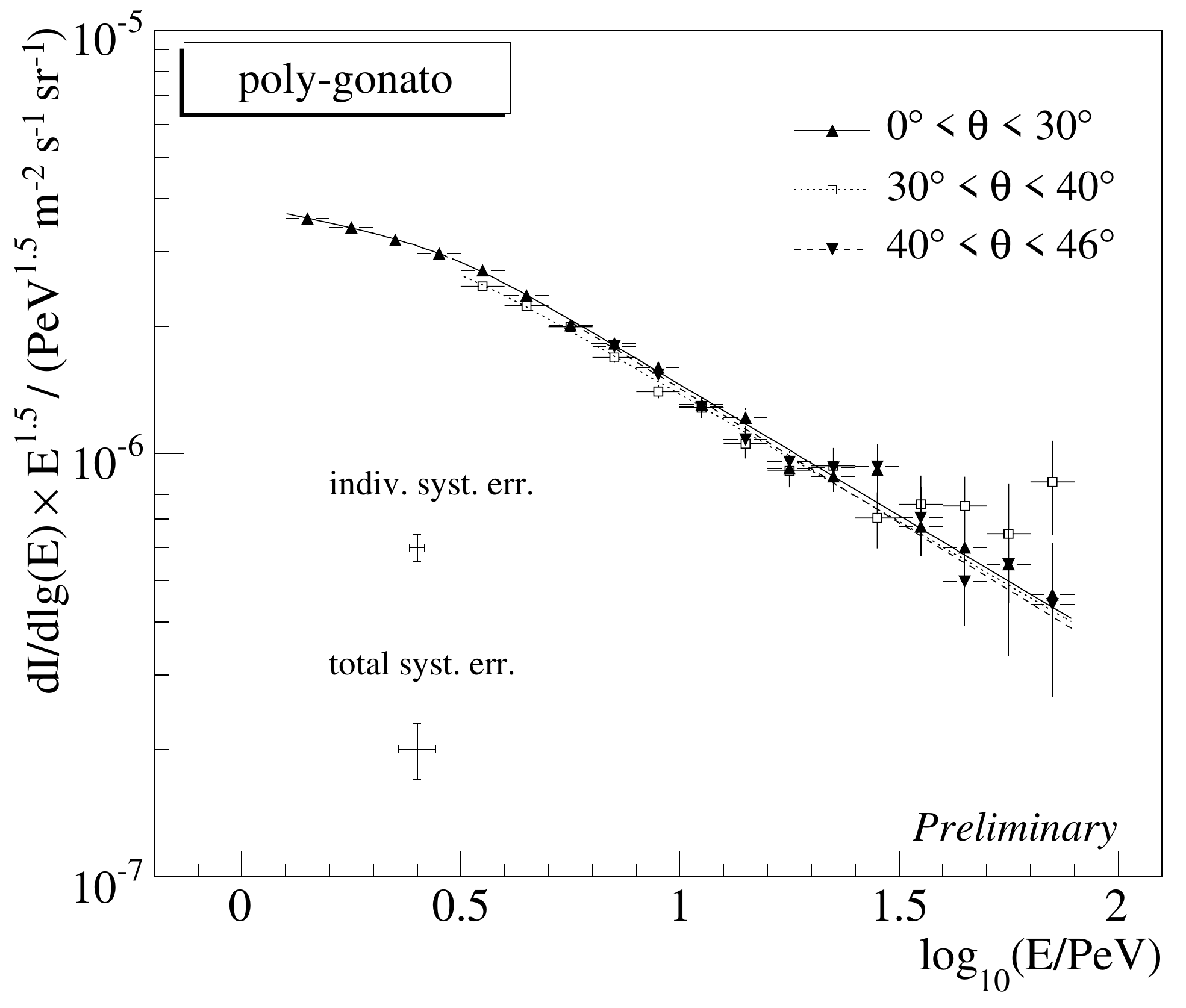} \label{poly}}
\hfil
{\includegraphics[width=0.47\columnwidth]{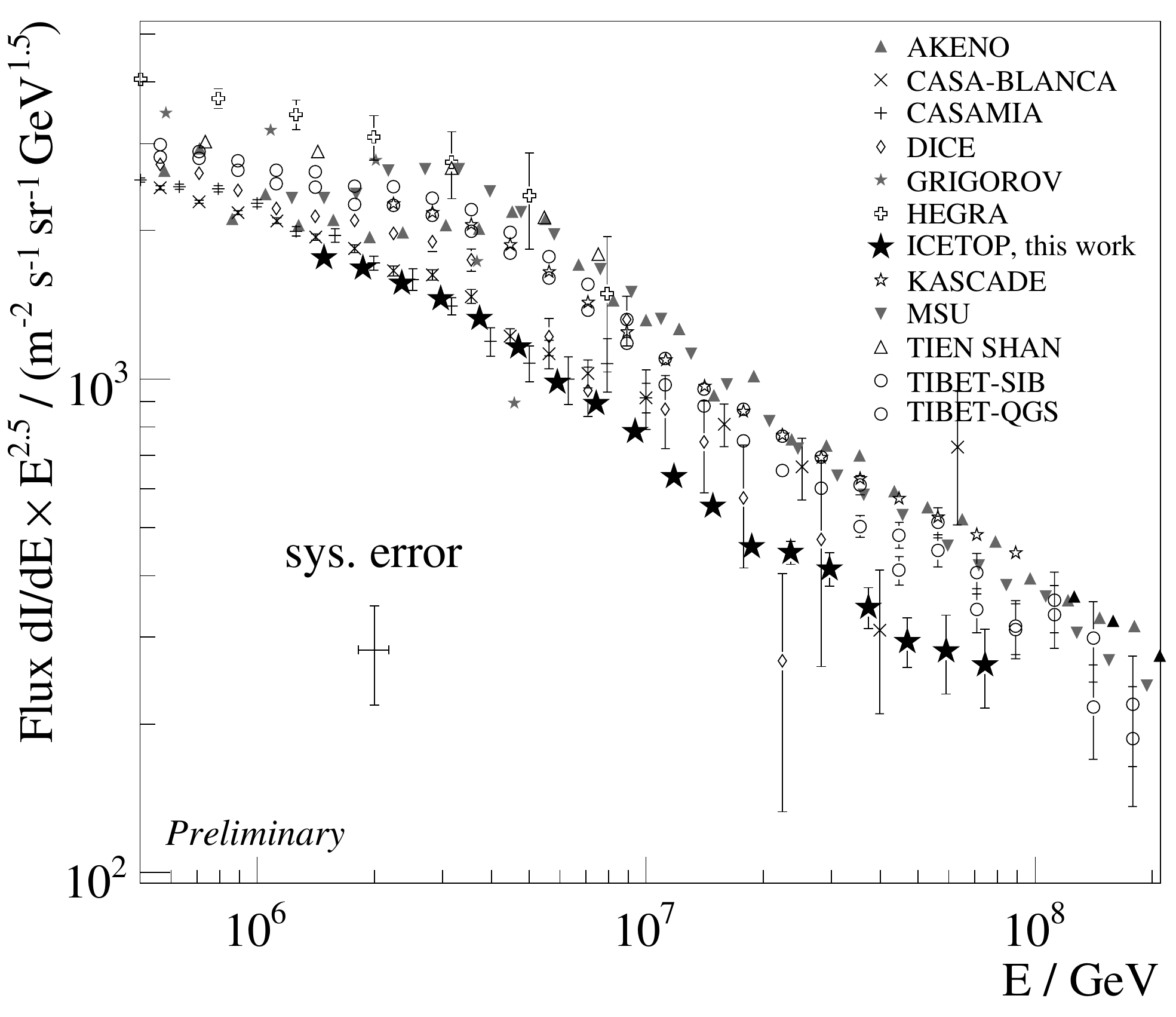}}}
\vspace{-2mm}
\caption{Left: Same as Fig.\,\ref{unfolding_pFe} for the poly-gonato model. Right: Preliminary energy spectrum from $1-80\eh{PeV}$ (scaled by $E^{2.5}$) measured by IceTop in August 2007 compared to results from other experiments. The error bars of the single points represent statistical errors. The total systematic error is displayed on the lower left.}
\label{unfolding_PGM}
\end{figure}

\section{ANALYSIS OF THE ENERGY SPECTRUM}
The air shower data recorded in August 2007 were divided into three zenith
bands, about equidistant in $\sec\theta$: $\Omega_0 = [0\dg,30\dg]$,
$\Omega_1 = [30\dg,40\dg]$ and $\Omega_2 =[40\dg,46\dg]$ (for details of the analysis see \cite{klepser_thesis}). For each of these angluar ranges, a proton and an iron response
matrix were simulated. In addition, two mixed composition
response matrices were calculated. One is a two-component
mixture of protons and iron \cite{two_components}; the iron fraction increases from 34\% at 1\,PeV to 80\% at 100\,PeV. The other one is
a 5-component implementation of the poly-gonato model that
was found optimal in \cite{hoerandel_ontheknee}. Here, the elements above helium increase from 40\% to 98\% in the same range.

Using the response matrices, the reconstructed energy spectrum was unfolded to yield the true energy spectrum. 
As can be seen in Fig.\,\ref{unfolding_pFe}, the resulting spectra assuming pure proton or pure iron primaries are not the same for different zenith angular ranges as expected for an isotropic flux, and they have an opposite ordering for protons and iron. Furthrmore, the proton spectra diverge towards higher energies, whereas the iron spectra converge. This suggests that the response matrix needed to reproduce an isotropic flux must be generated assuming a mixed composition with a mean mass increasing with energy. In fact, the spectra obtained with the poly-gonato model (Fig.\,\ref{unfolding_PGM}, left) as well as with the two-component model (not shown) do agree much better. 

The right plot in Fig.\,\ref{unfolding_PGM} shows the preliminary spectrum from
IceTop, assuming the 5-component poly-gonato composition
model along with several energy spectra from other experiments. The two-component model yields almost the same result.
The systematic uncertainty indicated in the plot is $10-11\prc$, slightly depending on energy, 
with the main contributions coming from the muon calibration ($7\prc$ in $E$) and some imperfections in the simulation.  These uncertainties can be appreciably reduced in the future. 
In the CORSIKA shower simulation~\cite{corsika} two high-energy interaction
models were tested up to now, namely SYBILL2.1 \cite{sibyll} and
QGSJet01.c \cite{qgsjet}. The  energy assignments obtained using these  models differed by less than $1\prc$; the small difference is probably due to the low muon content of the IceTop signal. 

The spectrum has the knee feature at
$3.1\pm0.3\,(\mr{stat.})\pm0.3\,(\mr{sys.})\eh{PeV}$. The power index changes
from $\gamma_1=-2.71\pm0.07\,(\mr{stat.})$ to
$\gamma_2=-3.110\pm0.014\,(\mr{stat.})$. The preliminary estimate of the systematic uncertainty
of the slopes is $0.08$. The absolute
flux is below most of the other spectra. Taking into
account the systematic error of our and the other measurements,
however, the deviation corresponds to no more than about
$1.5\,\sigma_{\mr{sys.}}$.

\section{SUMMARY AND OUTLOOK}
The first (preliminary) analysis of data taken with the 2007 configuration of the IceTop air shower array in the energy range from 1 to 80 PeV shows the expected change of slope in the energy spectrum and a mass composition consistent with mixed composition models such as the poly-gonato model. It was found in this analysis that the zenith angular dependence of the depth of the shower maximum can be exploited to analyse the mass composition. For such an analysis the high altitude of the detector, being close to the shower maximum, is very favourable.

In the future, with increasing detector size, the energy range, the systematic uncertainties, and the composition sensitivity will be appreciably improved. The mass composition can be particularly well derived by exploiting coincidences of IceTop with the in-ice detector yielding the electromagnetic versus the high-energy muon content of the showers. This is the special feature and strength of the IceCube experiment. In addition, other methods to extract composition sensitive quantities from the data are investigated: Muons with large transverse momenta with respect to the muon bundle axis are expected to be more abundant for lighter primaries. It was also found that muons can be counted with the surface detector alone, though probably restricted to the outer shower regions where the average tank signals drop below the average muon signal. Combining these different composition measurement methods allows, besides a more accurate composition determination, a scrutiny of the shower models, which is particularly important as the models cannot be tested directly.

\end{document}